\documentclass[]{amsart}

\usepackage{wrapfig}
\usepackage{amsmath,amssymb, amsfonts, mathtools}
\usepackage{verbatim}

\usepackage{newtxtext,newtxmath}       
\usepackage{helvet}         
\usepackage{courier}        
\usepackage{graphicx}        
\usepackage{multicol}        
\usepackage[bottom]{footmisc}
\usepackage{amsmath,amssymb, amsfonts, mathtools}
\usepackage{verbatim, booktabs, geometry}
\usepackage{bbm, bm}
\usepackage{algpseudocode, algorithm}
\usepackage{wrapfig, caption}
\usepackage{chngcntr}
\usepackage{apptools}
\usepackage{xcolor, color}
\usepackage{appendix}
\usepackage[]{hyperref} %
\usepackage{url}
\usepackage{lineno}
\usepackage{diagbox}

\usepackage{subcaption}
\usepackage{multirow}

\usepackage{lipsum, enumitem}
\usepackage{epstopdf}
\usepackage{varwidth}
\usepackage[export]{adjustbox}
\usepackage[edges]{forest}

\newcommand{\R}{\mathbb{R}}
\newcommand{\E}{\mathbb{E}}

\newcommand{\cN}{\mathcal{N}}

\DeclareMathOperator{\Var}{\mathrm{Var}}

\newcommand{\dpc}{d_p^c}
\newcommand{\given}{\,|\,}

\numberwithin{equation}{section}

\ifpdf
\DeclareGraphicsExtensions{.eps,.pdf,.png,.jpg}
\else
\DeclareGraphicsExtensions{.eps}
\fi

\let\oldState\State
\newcommand*{\stopnumbering}{%
	\let\olditem\item
	\renewcommand{\item}[1][]{\olditem[]}%
	\let\State\Statex}
\newcommand*{\resumenumbering}{%
	\let\item\olditem
	\let\State\oldState}

\makeatletter
\renewcommand{\ALG@name}{Algorithm}
\makeatother

\graphicspath{{./images/}}
\usepackage[noabbrev]{cleveref}

\makeatletter
\def\@setthanks{\vspace{-\baselineskip}\def\thanks##1{\@par##1\@addpunct.}\thankses}
\makeatother

\begin{document}

\title{Materials Fingerprinting Classification}
\author{Adam Spannaus$^1$}
\thanks{This manuscript has been authored by UT-Battelle, LLC under Contract No. DE-AC05-00OR22725 
	with the U.S. Department of Energy. The United States Government retains and the publisher, 
	by accepting the article for publication, acknowledges that the United States Government 
	retains a non-exclusive, paid-up, irrevocable,world-wide license to publish or reproduce 
	the published form of this manuscript, or allow others to do so, for United States 
	Government purposes. The Department of Energy will provide public access to these 
	results of federally sponsored research in accordance with the DOE Public 
	Access Plan (http://energy.gov/downloads/doe-public-access-plan).}
\address[1]{Oak Ridge National Laboratory, Oak Ridge, TN 37830}
\author{Kody J. H. Law$^2$}
\address[2]{School of Mathematics, University of Manchester, Manchester, UK}
\author{Piotr Luszczek$^3$}
\address[3]{Innovative Computing Laboratory, University of Tennessee, Knoxville, TN 37996}
\author{Farzana Nasrin$^4$}
\address[4]{Department of Mathematics, University of Hawaii at Manoa, Honolulu, HI 96822}
\author{Cassie Putman Micucci$^5$}
\address[5]{Eastman Chemical Company, Kingsport, TN 37662}
\author{Peter K. Liaw$^6$}
\address[6]{Department of Materials Science and Engineering, University of Tennessee, Knoxville, TN 37996}
\author{Louis J. Santodonato$^7$}
\address[7]{Advanced Research Systems, Inc., Macungie, PA 18062}
\author{David  J. Keffer$^6$}
\address[6]{Department of Materials Science and Engineering, University of Tennessee, Knoxville, TN 37996}
    \email[Corresponding author]{dkeffer@utk.edu}
\author{Vasileios Maroulas$^{8,*}$}
\address[8]{Department of Mathematics, University of Tennessee, Knoxville, TN 37996}
    \email[Corresponding author]{vmaroula@utk.edu}

\renewcommand{\shortauthors}{Spannaus et al.}
\begin{abstract}
    Significant progress in many classes of materials could be made with the availability of experimentally-derived large datasets composed of atomic identities and three-dimensional coordinates. Methods for visualizing the local atomic structure, such as atom probe tomography (APT), which routinely generate datasets comprised of millions of atoms, are an important step in realizing this goal. However, state-of-the-art APT instruments generate noisy and sparse datasets that provide information about elemental type, but obscure atomic structures, thus limiting their subsequent value for materials discovery.  The application of a materials fingerprinting process, a machine learning algorithm coupled with topological data analysis, provides an avenue by which here-to-fore unprecedented structural information can be extracted from an APT dataset. As a proof of concept, the material fingerprint is applied to high-entropy alloy APT datasets containing body-centered cubic (BCC) and face-centered cubic (FCC) crystal structures.  A local atomic configuration centered on an arbitrary atom is assigned a topological descriptor, with which it can be characterized as a BCC or FCC lattice with near perfect accuracy, despite the inherent noise in the dataset.  This successful identification of a fingerprint is a crucial first step in the development of algorithms which can extract more nuanced information, such as chemical ordering, from existing datasets of complex materials.
\end{abstract}


\maketitle

\section{Introduction}

Recent advancements in computing and contemporary machine-learning technologies have yielded new paradigms in computational materials science that are accelerating the pace of materials research and discovery~\cite{butler2018machine,curtarolo2013high,islam2018machine,zhou2018learning,ziletti2018insightful}.
For example, researchers have used a 
neural network to predict materials properties, clustering them
into groups consistent with those found on the periodic table~\cite{zhou2018learning} and
data-driven
materials design is an area now available to researchers
due to advances in machine-learning algorithms and computational materials science databases
~\cite{islam2018machine,agrawal2016perspective,NOMAD}. These developments in computational materials science 
have led researchers to begin exploring structure-property relationships
for disordered materials,
such as entropy-stabilized oxides and high-entropy alloys (HEAs)~\cite{rost2015entropy,zhang2014microstructures}.
Considering the number of atomic configurations in a disordered crystal structure, such
as those found in HEAs~\cite{zhang2008solid}, the number of 
possible atomic 
combinations of even a single unit cell,
the smallest collection and ordering of atoms from which an entire material can be built, quickly becomes computationally
intractable for existing algorithms~\cite{butler2018machine}.
In the present work, we propose an automated machine learning methodology for 
determining the lattice structure of a noisy and sparse materials dataset, e.g., the type 
retrieved from atom probe tomography (APT) experiments, 
for materials with disordered lattice structures, such as HEAs.

One of the fundamental properties of a crystalline material is the structure of its 
unit cell. Indeed, knowledge of the chemical ordering and geometric arrangement
 of the atoms of any 
material is essential for developing predictive structure-property relationships. 
As materials become 
more complex and the ordering of atoms amongst lattice sites becomes increasingly disordered, 
such as is the case with HEAs~\cite{yeh2004nanostructured}, these structure-property
relationships have yet to be developed.
Indeed, the high-configurational entropy of HEAs 
yields a distribution of lattice parameters and cell compositions, as 
opposed to a single unit cell and lattice constant found in more 
traditional materials.

For many classes of materials,
the lattice structure is either well-known, e.g., sodium chloride (salt) is body-centered cubic, or it can be discovered via X-ray diffraction (XRD) or neutron scattering
techniques~\cite{santodonato2015deviation}. 
XRD is a routine technique for the determination of crystal structures 
of metals, ceramics, and other crystalline materials. 
These techniques do not yield atomic level elemental distinctions or
resolve local lattice distortions on a scale of less than 10\AA\ \cite{santodonato2015deviation},
which are crucial to researchers working with highly-disordered materials, such as HEAs. Moreover,
XRD cannot provide the correlation between atom identity and position in a material.  
This chemical ordering of atoms is essential to developing predictive
relationships between the composition of an HEA and its properties.

High entropy alloys are a relatively new class of metallic alloys, first synthesized in
the mid 2000's by~\cite{yeh2004nanostructured}. As 
defined by~\cite{yeh2015physical}, 
HEAs are composed of at least five atomic elements, 
each with an atomic concentration between 5\% and 35\%.
These novel alloys have remarkable
properties, such as: corrosion resistance~\cite{zhang2014microstructures,shi2017corrosion},
increased strength 
at extreme temperatures, ductility~\cite{gludovatz2014fracture,lei2018enhanced,li2016metastable},
increased levels of elasticity~\cite{tsai2014high}, 
strong fatigue and 
fracture resistance~\cite{gludovatz2014fracture,hemphill2012fatigue,tang2015fatigue},
and
enhanced electrical conductivity~\cite{guo2017robust,kovzelj2014discovery}.
HEAs are amenable to the APT analysis as the process is able to 
recover elemental type in addition to approximate the lattice
sites in a material where the atoms sit. 

An experimental process that unambiguously 
determines the position, identity of each atom, and structure of a material is currently nonexistent~\cite{butler2018machine,moody2011lattice}.
Indeed,
quantification of different lattice parameters and unit-cell compositions have 
not previously been reported due to data quality issues inherent to 
APT~\cite{kelly2013atomic,miller2012future}.
While these experiments are able to discern elemental types
at a high resolution, the process has two drawbacks, \emph{(i) sparsity}: 
empirically, approximately 65\% of the atoms from a sample are not registered 
by the detector~\cite{santodonato2015deviation}; and \emph{(ii) noise}: the atoms
that are observed have their spatial coordinates corrupted 
by experimental noise~\cite{miller2012future}.
As
noted by~\cite{miller2012future}, the spatial resolution of the APT process is up to 3\AA\ 
(0.3 nm) in the $xy$-horizontal plane, which is approximately the length of 
an unit cell. 
This experimental noise has a two-fold impact on the data retrieved by
a typical experiment. First, the noise prevents materials science researchers from extracting elemental atomic distributions, which
are essential for developing the necessary interaction potentials for
molecular dynamics simulations. Secondly,
the experimental noise is significant enough to make atoms that are first neighbors 
in an atomic neighborhood, i.e., those atoms which occupy adjacent lattice sites,
appear as second or third neighbors and vice versa~\cite{miller2012future}.
Furthermore, the experimental noise is only one source of 
distortion to the lattice structure. HEAs exhibit local lattice deformations
due to the random distribution of atoms throughout the 
material and 
atoms of differing size sitting at adjacent lattice points~\cite{zhang2008solid}. 

This deformation of the local crystal structure makes any
determination of the lattice a
challenging problem for any symmetry-based algorithm, such
as~\cite{hicks2018aflow,honeycutt1987molecular,larsen2016robust}. 
The field of atom probe crystallography has emerged in recent 
years~\cite{moody2011lattice,gault2012atom} and
existing methodologies in this area
seek to discover local structures when the global structure is known
\emph{a priori}. In the case of HEAs, the global lattice structure is unknown and must be
discovered. Indeed, drawing correct conclusions about the material's crystal structure is
virtually impossible from the APT analysis using current techniques~\cite{miller2012future}.

A recent method relying on a convolutional neural network~\cite{ziletti2018insightful}
classified synthetic crystal structures that are either noisy or sparse by creating a
diffraction image from a lattice structure and using this image as input data for the neural
network. The authors of~\cite{ziletti2018insightful} claim that their methodology could be applied 
to data with both experimental noise and significant levels of sparsity, as is 
	typically retrieved by APT experiments, but without showcasing any such instances.
Briefly, diffraction images are diffraction patterns generated by simulating the
results of an X-ray diffraction experiment. In particular, they create the interference pattern
that is generated when a series of waves encounter a crystal lattice and either pass through
unobstructed or encounter an atom and bend around the atom. 

\begin{figure}
\centering
        \includegraphics[width=\textwidth]{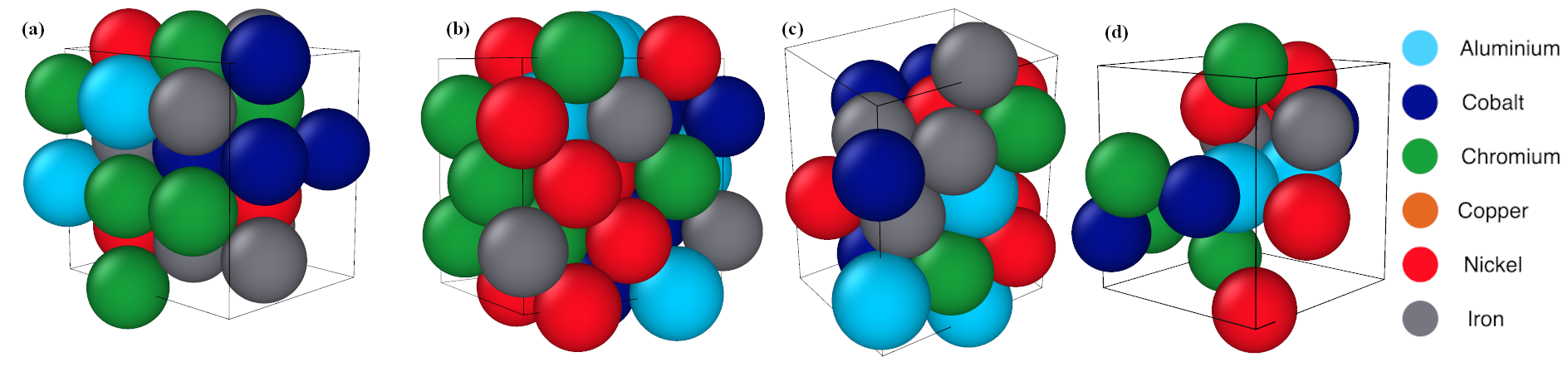}
    \caption{Examples of the lattice structures that 
    		we consider viewed with the visualization software Ovito~\cite{ovito}
    		which uses empirical atomic radii in its visualizations. We consider three different
    		crystals: 
    		(a) body-centered cubic (BCC), (b) face-centered cubic (FCC), and 
    		(c) hexagonal close packed (HCP) lattices showing their similarities and differences 
    		with complete, noiseless data. The FCC and 
    		HCP structures have only a subtle difference in their geometry. The HCP structure
    		forms an identifying parallelogram (c), whereas the FCC 
    		forms a square (b) when all atoms within a radius of the center atom are
    		collected. 
    		(d) Example of an FCC structure retrieved from an APT analysis of the HEA Al$_{0.3}$CoCrFeNi~\cite{diao2019novel}
    		demonstrating the sparsity and atomic displacements due to the resolution of APT process.
    		The noise and sparsity from the APT process obscures this difference between the FCC and HCP 
    		structures.}
    \label{fig:apt_noise}
\end{figure}

Here we propose a machine-learning approach, a materials fingerprint, to classify the crystal structure of a material by looking at local atomic neighborhoods through the lens of topological data analysis (TDA). TDA is a field that uses topological features within data for machine learning tasks~\cite{Marchese2018,marchese2017kappa,nasrin2019bayesian}. 
It has found other applications in materials 
science, such as the characterization of amorphous solids~\cite{hiraoka2016hierarchical}, equilibrium phase transitions~\cite{donato2016persistent}, and similarity of pore-geometry in nanomaterials~\cite{lee2017quantifying}.
Our motivation is to encode the geometric peculiarities of HEAs by
considering atomic positions within a neighborhood and looking at the neighborhood's topology. 
Key differences between atomic neighborhoods are encoded in the empty space, e.g., holes and voids, between atoms, as well as clusters of atoms in the neighborhood. 
These identifying topological features of an atomic neighborhood can be calculated through the concept of homology, 
which is the mathematical study of `holes' in different dimensions and 
differentiate the shape and structure of the neighborhoods.
Extracting this homological information from each atomic neighborhood, we can distinguish between 
the different lattice structures that we consider; \cref{fig:apt_noise} shows idealized 
	versions of these crystal structures. 
	A typical lattice retrieved from an APT experiment is in~\cref{fig:apt_noise}(d).

Using these topologically-derived features, we are able to classify the crystal structure 
of HEAs from the APT data with accuracy approaching 100\%.
To test the robustness of our proposed method, we combine levels of sparsity and noise on synthetic data and find our 
method accurately classifies the crystal structure. Our novel methodology couples the power of topological data
	analysis to extract the intrinsic topology of these
	crystal lattices with a machine learning classification scheme
	to differentiate between lattice structures and classify them with a high degree of precision.

The outline of this paper is as follows. In~\Cref{sect:APT} we describe the APT experimental process and 
	the details related to the analysis of the HEAs that we consider. \Cref{sect:methods} provides details of the
	 classification model for recognizing crystal structures. Numerical
	results are presented in~\cref{sect:results} and we conclude with discussion in~\cref{sect:conclusion}.


\section{Atom Probe Tomography}\label{sect:APT}

In this section we discuss the APT experimental process and the postprocessing
employed to create the data. Furthermore, we discuss the resulting data and its
characteristics.

\subsection{APT Process}\label{sect:apt_process}

APT was conducted using a Local Electrode Atom Probe (LEAP) 4000 XHR instrument at the Center 
		for Nanophase Materials Sciences of the Oak Ridge National 
		Laboratory~\cite{diao2019novel,guo2016atom}.
	The process
	systematically evaporates ions from a specimen’s hemispherical surface using 
	voltage or laser pulses. A position sensitive detector collects the ions, and 
	the timing between the pulse and detection events gives the time-of-flight, which 
	identifies each species based on unique mass-to-charge ratios. A reconstruction 
	algorithm is used to create a tomographic dataset from the $x$, $y$ detector data 
	and the sequence of detection gives the $z$-dimension in the reconstruction. 
	Sample specimens for APT experiments 
	are typically sharp, conical tips with a maximum 
	diameter of less than 100 nm and a length of several 
	hundred nanometers typically. Thus all APT experiments investigate nanoscale structures and 
	samples that contain nanoparticles embedded in a matrix can be examined as well as  
	layered heterostructures.

\subsection{APT Data}\label{sect:apt_data}

For our problem, the data consists of spatial coordinates of approximately $10^8$ atoms with elemental type~\cite{miller2012future}, constituting a highly-disordered metallic alloy that is 
composed of BCC or FCC lattice structures. 
The sample~\cite{santodonato2015deviation} was chosen because it has been previously well-characterized. 
This alloy consists of three phases, a Cu-rich FCC phase, an Fe-Cr rich BCC phase, and a remaining phase that incorporates all six elements, though the proportions of Cu, Fe, and Cr are depleted due to accumulation in the other phases.  Importantly all three phases are present in the APT sample.  When viewing the entire data set with atoms identified by color, some nanoscale information is immediately evident. 
The eye perceives elemental segregation of the Cu-rich and Fe-Cr rich phases into nanoscale domains. The orange copper-rich area is especially evident, as seen in figure 2(a).
However, one cannot infer any meaningful structure at a finer scale when viewing the entire dataset from a typical APT experiment and further analysis requires that individual atomic neighborhoods be extracted from the larger sample. 
Viewing each neighborhood individually, figure 2(b), we can see that they contain a wealth of information about the shape of the material under investigation, despite the noise and sparsity present in a typical APT experiment.

\section{Methods}\label{sect:methods}

In this section we give the mathematical background necessary for our method,
detailed introductions can be found in ~\cite{edelsbrunner2008persistent,Ed2010}.

\subsection{Topological Data Analysis}\label{sect:TDA}

To extract the salient topological information from the atomic neighborhoods, we turn to topological data analysis, particularly persistent homology.
Persistent homology describes connectedness and void space present within an object and
allows one to infer global properties of space from local information~\cite{kaczynski2006computational}.
Instead of considering only clusters of atoms, homology also 
incorporates information about the regions enclosed by the atoms.
This approach yields topological features of the data in different homological dimensions. 
In the case of these atomic neighborhoods created by APT experiments, 
$0-$dim homological features are connected components,
$1-$dim homological features are holes, and $2-$dim homological features are voids, 2-dim holes, i.e., the space enclosed by a sphere.

\begin{figure}[t]
	\centering
	\includegraphics[width=\textwidth]{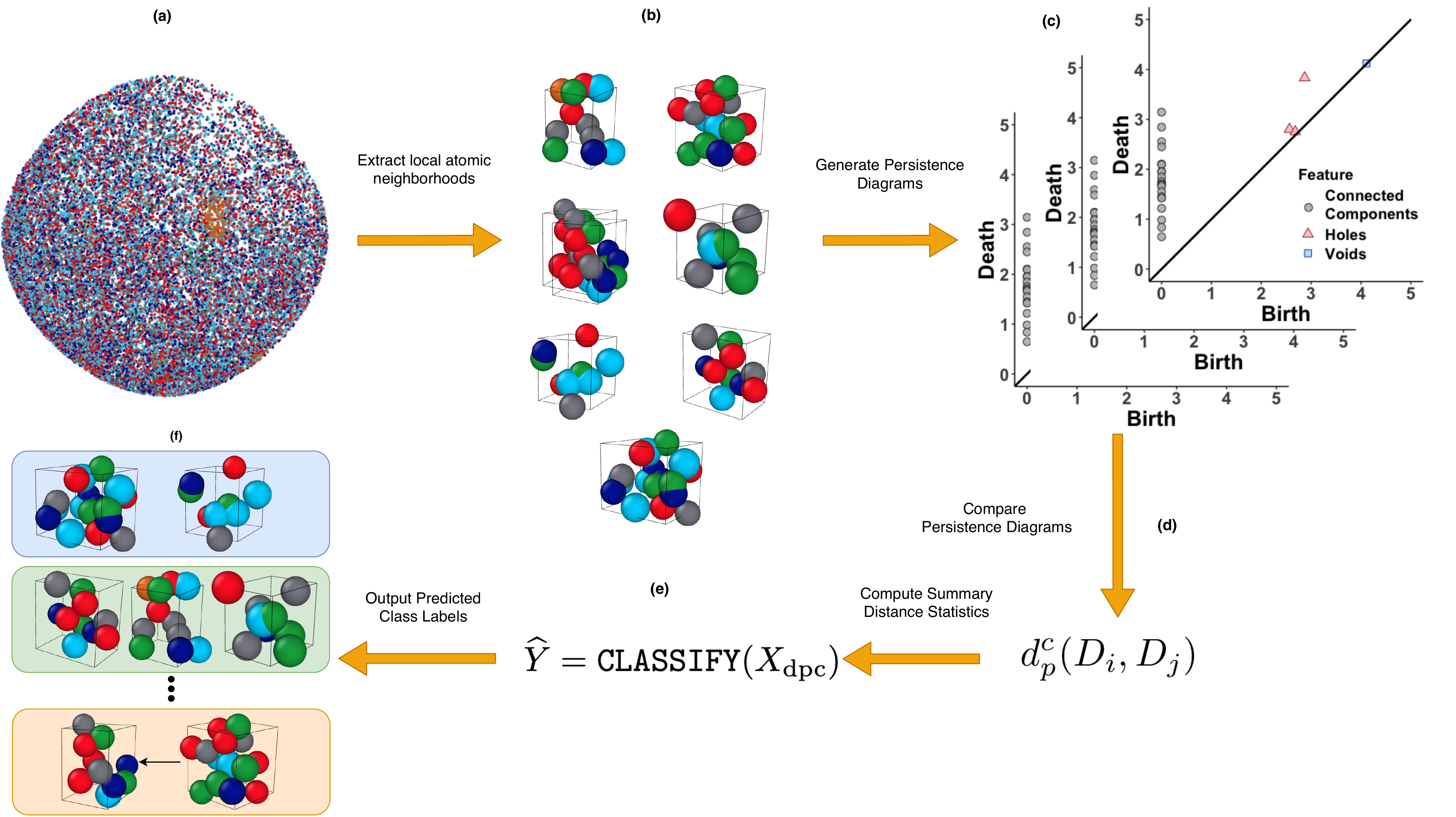}
	\caption{Flowchart of the materials fingerprinting methodology. (a) The APT data is processed as outlined in~\cref{sect:apt_process}. (b) Individual atomic neighborhoods are extracted from an APT dataset as described in~\cref{sect:apt_data}.
		 (c) We create a collection of persistence diagrams, each diagram associated with an atomic neighborhood, 
		 as explained in~\cref{sect:TDA}. (d) Similarity metrics between these persistence diagmras
		 are computed via the $\dpc$-distance as defined in~\cref{eqn:dpc}. (e) We create a feature matrix composed of the summary statistics of these distances, which is used as input in~\cref{alg:our_scheme} to classify the persistence diagrams. 
		 (f) Output from~\cref{alg:our_scheme} classifying the structures under investigation, \cref{sect:methods}.}
	\label{fig:fingerprint}
\end{figure}

\begin{figure}[t]
\centering
        \includegraphics[width=\textwidth]{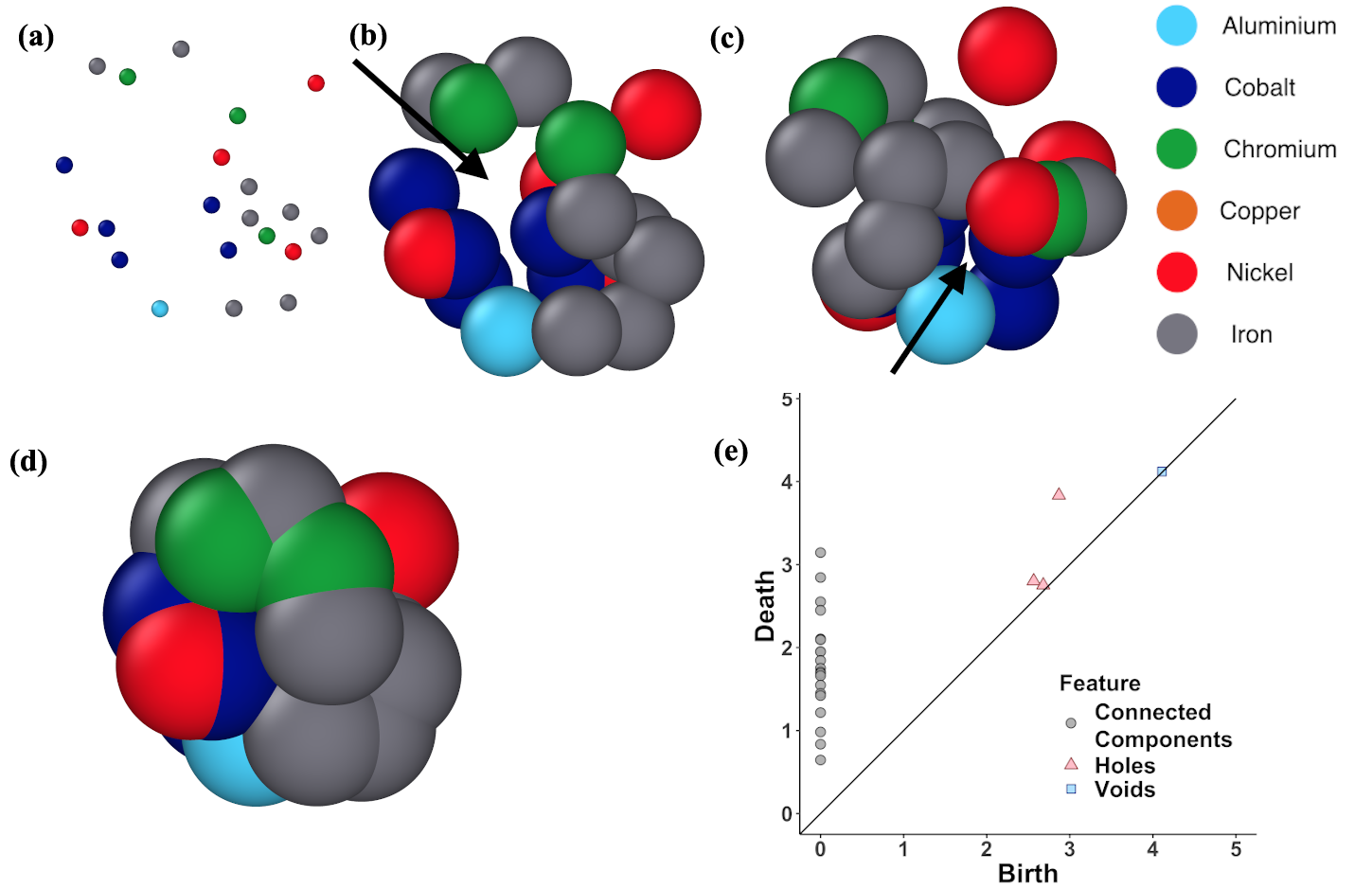}
        \caption{Atomic neighborhood from an APT experiment~\cite{santodonato2015deviation} with the alloy Al$_{1.3}$CoCrCuFeNi. The atomic type is illustrated by the color, and is visualized with~\cite{ovito}. (a) Shows each atom in a neighborhood
        		 as a point cloud in $\R^3$. We begin by drawing a radius centered at 
        	each atom. As the radius of these spheres increases in (b), a $1-$dim hole forms in the atomic structure. Increasing the radii further, in (c) the formation of a $2-$dim hole, a void, is evident. Continuing to increase the radii, in (d) the radii have increased such that all atoms form one cluster.  The persistence diagram for this structure is shown in (e). In the persistence diagram, the birth and death axes denote the emergence or disappearance of topological features as the radii of the spheres centered on
        each atom increase and start to intersect.} 
	\label{fig:homology}
\end{figure}

To study the persistent homology of atomic structures extracted by HEAs,
such as the atomic neighborhoods in~\cref{fig:fingerprint}(b), we create spheres of increasing radii around each atom in a neighborhood, detect when homological features emerge and disappear, and record these radii in a persistence 
diagram, see figures \ref{fig:fingerprint}(c) and \ref{fig:homology}(e). Taking the atoms' spatial positions in the $xyz$-coordinate system
recovered by the APT experimental process,
 we begin by considering a sphere of radius $\epsilon$ centered at each atom, see~\cref{fig:homology}(a). 
The algorithm starts at radius $\epsilon=0$ and this is 
the reason why all points start at 0 in the persistence
diagram associated with clusters and connected components (grey circles in figures 2(c) and 3(e)).
Indeed, all atoms within a structure are initially treated as different clusters. 
Increasing the radii, the algorithm starts clustering atoms 
together by examining if their spheres intersect at a certain radius. If they do,
the these atoms form a cluster and that signifies the `death' of the members of clusters
as being considered separately. Meanwhile, as spheres grow holes and voids (2-dim holes) are created, see
figures 3(b) and 3(c). By the same token, these holes and voids
get filled in due to increasing the radii, and are represented
in a persistence diagram by their death time (radius-wise). Indeed, such topological features
are recorded in a persistence diagram using a different label (color). 
Eventually, at some radius, all spheres will intersect, which means that all atoms belong to the same
cluster and any hole or void has been covered. This yields the end of the algorithm for creating a persistence diagram. These homological features summarized in a persistence diagram
capture information about the shape of the neighborhood itself. This type of multiscale analysis is key to bypassing the noise and sparsity present in the data and to extract meaningful details about the configuration of each neighborhood. For example, 
 the corresponding diagram for the atomic neighborhood in~\cref{fig:homology}(a) is shown
 in~\cref{fig:homology}(e). 
 The persistence diagram encodes information about the structure of each neighborhood by providing insight about the number of atoms, the size and distance among atoms, possible configuration of the faces, and $3-$dimensional structure.  The persistence diagram then functions as a proxy for 
 the APT data by reducing an atomic neighborhood to its most pertinent qualities.


\begin{figure}[]
	\centering
	\includegraphics[width=\textwidth]{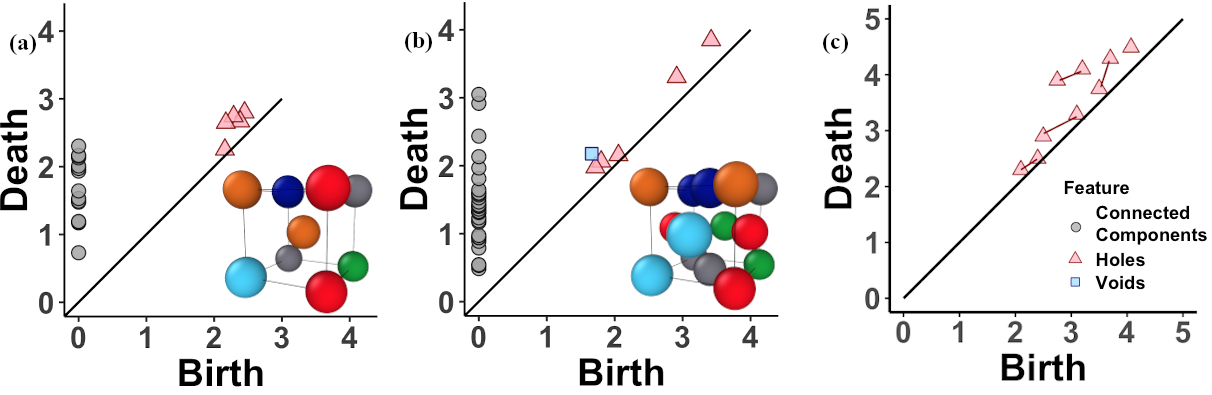}
	\caption{Sample persistence diagrams of a material from the APT analysis of the alloys Al$_{1.3}$CoCrCuFeNi and Al$_{0.3}$CoCrFeNi for two of the lattice types considered here: BCC (a) and FCC (b), respectively~\cite{santodonato2015deviation,diao2019novel}. Notice the distinguishing $2-$dim feature, the blue square, in the diagram derived from an FCC lattice.  Additionally, the diagram generated from the BCC structure has fewer $0-$dim features.
	(c) The $d_p^c$ metric computes the distance between two persistence diagrams generated by atomic neighborhoods, both containing 1-dim features, denoted by the red triangles. 
		The $d_p^c$ metric measures the distance between the diagrams by first finding the best matching between points, given by the lines between the triangles. 
		 Any unmatched points, e.g., the 
		remaining triangle, are then penalized by the constant term $c$. The birth and death axes denote the emergence or disappearance of topological features, as a function of distance between atoms in a neighborhood.}
	\label{fig:materials_pd}
\end{figure}

As the extracted persistence diagrams generated by APT experiments summarize the shape peculiarities of each atomic neighborhood, different types of 
lattice structures yield persistence diagrams with various identifying 
features~\cite{maroulas2019stable}. Indeed, examining the homological features, 
we see the essential structural differences between crystal lattices 
in different dimensions. 
Consider~\cref{fig:materials_pd}, which displays the difference between persistence diagrams for BCC and FCC structures. From the viewpoint of topology, the inside of an FCC cell contains a void, whereas the BCC cell does not, 
thus yielding an important contrast. In the case of noiseless and complete data, the presence of a void separates the BCC and FCC cells when juxtaposing their crystal structures, as we see in the insets of~\cref{fig:materials_pd} (a,b). 
The persistence diagrams capture differences in (i) the number of neighbors (8 for BCC and 12 for FCC), (ii) the spacing between neighbors, i.e., density, and (iii) the arrangement of neighbors.

\subsection{Persistence Diagram Similarity Metric}\label{sect:dpc_dist}

Different crystal structures produce different size point clouds~\cite{maroulas2019stable}.
To properly account for differences in the number of points when comparing two persistence diagrams, we employ the $d_p^c$  distance, introduced in~\cite{Marchese2018}. 
 For a given configuration, the persistence diagram can be compared to a reference persistence diagram via a similarity metric,
for BCC and FCC structures as an example.
Suppose $D_1= \{ d^{1}_1, \ldots, d^{1}_n \}$ and 
$D_2 = \{ d^{2}_1, \ldots, d^{2}_m \}$ are two persistence diagrams 
associated with two local
atomic neighborhoods such that $n\leq m$. 
Let $c > 0$ and $1 \leq p < \infty$ be fixed parameters. 
Then the $d_p^c$ distance between $D_1$ and $D_2$ is
\begin{equation}\label{eqn:dpc}
d_p^c(D_1, D_2) = 
\left( \frac{1}{m} \left( \min_{\pi \in \Pi_m} \sum_{i=1}^n \min(c,\|d^{1}_i-d^{2}_{\pi (i)} \|_{\infty})^p + c^p |m-n| \right) \right)^{\frac{1}{p}}
\end{equation}
where $\Pi_m$ is the set of permutations of $(1, \dots, m)$. If $n > m$, define $d_p^c (D_1, D_2) := d_p^c (D_2, D_1)$. 

 This distance matches points between the persistence diagrams being compared, and those that are unmatched are penalized by a regularization term $c$.
 \Cref{fig:materials_pd}(c) shows an example of how 
 	the distance between two persistence diagrams is computed. 
 	We first find the optimal matching, denoted by the red lines between triangles. This matching between points
 	corresponds to the summation term in the distance. If the matched
 	distance is greater than $c$, then we add $c$ to the matching distance,
 	otherwise, we add the distance between matched points.
 	 The unmatched 1-dim feature, denoted by the red triangle, is penalized
 	by the regularization term $c$ in the second part of the definition.
 In developing the materials fingerprint, we compare persistence diagrams with respect to 0, 1, and $2-$dim homological features, i.e., connected components, holes, and voids, employing this distance. We then compute summary statistics (mean, variance) from these distances to create features for the classification algorithm.

\subsection{Classification Model}\label{sect:classification_model}

 We write $D_i$ as the persistence diagram generated by atom positions
 in an atomic neighborhood retrieved by the APT experiment as seen in 
 \cref{fig:fingerprint}. Note that the number of atoms in
 a neighborhood is not constant, but varies between atomic neighborhoods
 in a sample. For the multiclass classification problem, we are interested in modeling 
	the conditional probability $\pi(Y=\ell\mid X)$ for
	a given input $X$, which encapsulates features of persistence diagrams
	and a class label $Y=\ell$.
	We write the classification model as a
	generalized additive regression
	model~\cite{friedman2000additive,HastieTrevor1990Gam}.
	 Choosing this type of 
	 model gives us the flexibility to let our data determine the correct 
	 functional form, as opposed to imposing a linear model as in 
	 traditional logistic regression.
	Accordingly, an $L$-class model is written
	\begin{align*}\label{eq:multiclass_regression}
	\log\left(\frac{\pi(Y = 1 \given X)}{\pi(Y=L \given X)}\right) &= \alpha_1 + F_1(X),\\
	\log\left(\frac{\pi(Y = 2 \given X)}{\pi(Y=L \given X)}\right) &= \alpha_2 + F_2(X),\\
	&\mathrel{\makebox[\widthof{=}]{\vdots}} \\
	\log\left(\frac{\pi(Y = L-1 \given X)}{\pi(Y=L \given X)}\right) &= \alpha_{L-1} + F_{L-1}(X),
	\end{align*}
	where $F_i (X) = \sum_{j=1}^P\,\alpha_j f_j(X)$ is a linear combination of smooth functions $f_j$.
	Here $\mathbf{X}\in\R^{N\times P}$ and
	$N = \sum_{i=1}^L\,N_i $ is such that for $1\leq i\leq N$ an arbitrary row 
	of $\mathbf{X}$ is
	\begin{equation}
	\mathbf{X}_i = {} (
	\E_{i,\lambda_1}^{0}, \E_{i,\lambda_1}^{1}, \E_{i,\lambda_1}^{2}, \Var_{i,\lambda_1}^{0}, \Var_{i,\lambda_1}^{1}, \Var_{i,\lambda_1}^{2},\dots,
	\E_{i,\lambda_L}^{0}, \E_{i,\lambda_L}^{1}, \E_{i,\lambda_L}^{2}, \Var_{i,\lambda_L}^{0}, \Var_{i,\lambda_L}^{1}, \Var_{i,\lambda_L}^{2}),\label{eqn:predictor}
	\end{equation}
	where $\E_{i,\lambda_\ell}^k =  \frac1{N_\ell} \sum_{j=1}^{N_\ell} \dpc(D_i^k,D_j^k)$
	and $\Var_{i, \lambda_\ell}^{k} = \frac1{N_\ell - 1}\sum_{j=1}^{N_\ell}\,(\dpc(D_i^k,D_j^k) - \E_{i,\lambda_\ell}^k)^2$
	are the mean and variance respectively of the $\dpc$ distance, \cref{eqn:dpc},
	between any diagram $D_i^k$
	and the collection of all persistence diagrams in the class 
	$\lambda_\ell\in\Lambda, 1\leq \ell\leq L$ and
	homological dimension $k=0,1,2$. 
	 The pseudocode for our algorithm is presented in~\cref{alg:our_scheme} and is visually
	represented in~\cref{fig:fingerprint}.
  
\begin{algorithm}[]
    \caption{Materials Fingerprinting}\label{alg:our_scheme}
    \begin{algorithmic}[1]
        \stopnumbering
        \State \textbf{Training Step}
        \resumenumbering
        \State Read in labeled data (training set) with $L$ classes and compute persistence diagrams in the training set $\mathcal{D}_{train}$, which has $N_\ell$ diagrams from the $\ell$th class, and set $N = \sum_{\ell=1}^L\,N_\ell$.
        \State Read in response vector 
        $Y = (1\cdot\mathbf{1}, \dots,\ell\cdot\mathbf{1},\dots,L\cdot\mathbf{1})^T$ where $\mathbf{1}$ is a 
        vector of 1's in $\R^{N_\ell}$.
        \For {$i = 1,\dots, N$}
        \State Compute feature matrix $\mathbf{X}$ according to equation~(\ref{eqn:predictor})
        \[
			\mathbf{X}_i = {} (
			\E_{i,\lambda_1}^{0}, \E_{i,\lambda_1}^{1}, \E_{i,\lambda_1}^{2}, \Var_{i,\lambda_1}^{0}, \Var_{i,\lambda_1}^{1}, \Var_{i,\lambda_1}^{2},\dots,
			\E_{i,\lambda_L}^{0}, \E_{i,\lambda_L}^{1}, \E_{i,\lambda_L}^{2}, \Var_{i,\lambda_L}^{0}, \Var_{i,\lambda_L}^{1}, \Var_{i,\lambda_L}^{2})
        \]
        where
        \begin{equation*}
        \E^k_{i,\lambda_\ell} = \frac1{N_\ell} \sum_{j=1}^{N_\ell} d_p^c(D_i^k,D_j^k),\,\,\,\,
        \Var_{i, \lambda_\ell}^{k} = \frac1{N_\ell-1}\sum_{j=1}^{N_\ell}\,(\dpc(D_i^k,D_j^k) - \E_{i,\lambda_\ell}^k)^2,
        \end{equation*}
        for $\lambda_\ell\in\Lambda, k\in\{0,1,2\}$.
        \EndFor
        
        \State $\mathcal{C}(\mathbf{X}) = \texttt{ADABOOST}(\mathbf{X}, Y)$\Comment Obtain a classification rule $\mathcal{C}$ from the AdaBoost ensemble classification algorithm
        \stopnumbering
        \State \textbf{Testing Step}
        \resumenumbering
        \State Read in unlabeled APT point cloud data and compute persistence diagrams $\mathcal{D}_{test} = \{\widehat{D}_j\}_{j=1}^J$.
        \For {$j = 1,\dots,J$}
        \State Compute
        \[
        \widehat{\mathbf{X}}_j = 
        (\widehat{\E}_{j,\lambda_1}^{0}, \widehat{\E}_{j,\lambda_1}^{1}, \widehat{\E}_{j,\lambda_1}^{2}, 
        \widehat{\Var}_{j,\lambda_1}^{0}, \widehat{\Var}_{j,\lambda_1}^{1}, \widehat{\Var}_{j,\lambda_1}^{2}, \dots,
        \widehat{\E}_{j,\lambda_L}^{0}, \widehat{\E}_{j,\lambda_L}^{1}, 
        \widehat{\E}_{j,\lambda_L}^{2}, \widehat{\Var}_{j,\lambda_L}^{0}, \widehat{\Var}_{j,\lambda_L}^{1}, \widehat{\Var}_{j,\lambda_L}^{2})
        \]
        where
        \begin{equation*}
        \widehat{\E}^k_{j,\lambda_\ell} = \frac1{N_\ell} \sum_{n=1}^{N_\ell} d_p^c(\widehat{D}_j^k,D_n^k),\,\,
        \widehat{\Var}_{j, \lambda_\ell}^{k} = \frac1{N_\ell - 1}\sum_{n=1}^{N_\ell}\,(\dpc(\widehat{D}_j^k,D_n^k) - \widehat{\E}_{j,\lambda_\ell}^k)^2,
        \end{equation*}
        for $\lambda_\ell\in\Lambda, k\in\{0,1,2\}$.
        \EndFor
        \stopnumbering
        \State \textbf{Classify unlabeled APT data}
        \resumenumbering
        \State $\widehat{Y} = \mathcal{C}(\widehat{\mathbf{X}})$\qquad\qquad\qquad\qquad\qquad\qquad \qquad$\triangleright$ Yields
        class labels for ${\mathcal{D}_{test}}$ as $\hat{Y}\in\{1,\dots,\ell,\dots,L\}^J$.
    \end{algorithmic}
\end{algorithm}

\subsection{Computational and Storage Considerations}

Computing entries of the feature matrix $\mathbf{X}$, equation~(\ref{eqn:predictor}),
requires computing the mean 
and variance 
of $\dpc$ distances with $k-$dim persistence homology, $k=0,1,2$. 
For example, in the case of binary classification between
BCC and FCC lattice types, with $N_1$ and $N_2$ neighborhoods respectively, 
for each BCC persistence diagram,
each $\mathbb{E}_{i,\lambda_1}^{k}$ computation requires $N_1$ steps and for FCC, it is $N_2$ steps.
Similarly, computing the variance accurately in a numerically stable fashion, e.g., when the size of the dataset is large
and the variance is small, each BCC diagram takes
$2\times N_1$ steps for the two pass algorithm~\cite{chan1983algorithms}. In total, each row of $\mathbf{X}$ has complexity
$\mathcal{O}_i(N_1,N_2)=9 \times \left( N_1+N_2 \right)$ and the entire feature
matrix ends up with quadratic complexity: $\mathcal{O}(N_1,N_2)=9 \times \left( N_1+N_2 \right)^2$.
With the atomic counts on the order of hundreds of thousands: $N_1,N_2 \approx
\mathcal{O}(10^5)$, the quadratic component clearly dominates with $10^{10}$
computational steps. Each of these steps requires the $\dpc$ distance computation 
given by~\cref{eqn:dpc},
which is computationally non-trivial for the majority of the
diagrams due to the identification of the optimal permutation
between the diagrams being compared. In order to reduce the total elapsed time of the computation, we used
over $1000$ x86 cores that ranged from Intel Westmere to Intel Skylake, ranging
in cores per socket from 8 to 36 with up to 72 cores per node. Additional
speedup of about $20\%$ came from porting the code for computing the 
feature matrix from Python to C. The python code is publicly 
available at 
\url{https://github.com/maroulaslab/Materials-Fingerprinting}.

\section{Numerical Experiments}\label{sect:results}

We present here the outcome of~\cref{alg:our_scheme} in 
both synthetic and real experimental data as well as provide a sensitivity
analysis. We first present results of our fingerprinting process in different scenarios with synthetic data
	to test the robustness of our method. 
	We consider synthetic APT data with various levels of 
	sparsity and additive Gaussian noise, $\cN(0, \sigma^2)$,
	as in real APT experimental data. 
In each of the experiments presented, we perform 10-fold cross validation on the entire 
dataset to control for overfitting of the model, randomly splitting the dataset into 
10 partitions.
For each partition, we create 
a classification rule from the other 9 partitions, and use the remaining one as a test set. 
Our accuracy, defined here as (1 - Misclassification rate), is recorded for each partition as it is
used as the test set. The reported accuracy rate is the mean accuracy over all 10 partitions.
The hyperparameters $c, p$ were set to the same values across all experiments, $c = (1, 0.05, 1.05)$
	and $p=2$,
	to provide a fair basis for comparison, and were selected by a grid search to provide 
	the highest accuracy score in the binary classification problem with 67\% missing data and
	$\cN(0,1)$ additive noise. A previous
	work~\cite{maroulas2019stable}, discusses the role of $c$ and choosing this parameter.

\subsection{Synthetic APT Data}

We first present results of our fingerprinting process in different scenarios with synthetic data
	to test the robustness of our method. 
	First, we test with combinations of noise and sparsity that we expect
to see in real APT data. Next, we examine the effect of class imbalances on the 
accuracy of our methodology
in the binary classification case of BCC and FCC materials. As a final experiment with synthetic data,
we repeat the scenario of varying the concentration between BCC and FCC structures, but augment 
the data set with a constant number of HCP lattice types. We observe the methodology
is robust against different levels of noise and sparsity in the case of the binary classification
problem. When the HCP structures are introduced into the dataset, the accuracy
decreases, due to the similarity of the FCC and HCP structures,
especially in
the presence of additive noise and sparsity that we consider. These 
results are presented in~\crefrange{tab:results}{tab:vary_prop_all3}.

%

\subsection{Sensitivity Analysis}\label{sect:sensitivity}

To understand the effect of different levels of noise and 
sparsity in the data, the materials fingerprint was applied to synthetic data having different levels of sparsity and noise, similar to those values found in real APT data. For each combination presented, we the dataset was composed
of 400 structures, split evenly between BCC and FCC types. We observe perfect accuracy
in the case of complete, noiseless data, as these lattice types differ in both their geometry and
atomic density. As the data becomes increasingly degraded, the accuracy correspondingly decreases, but
does not fall below 90\% in this analysis. \Cref{tab:results} summarizes these results.
We do observe a relative decrease in accuracy with 50\% sparsity and $\cN(0,0.75^2)$
added noise. We attribute this decrease to the choice of $c$ and $p$
for the distance computations. Indeed, for all the experiments presented
herein, we used the same values of $c$ and $p$. We may further optimize these
parameters to produce higher accuracy for each combination of noise and missing data
considered, at the risk of over-fitting for a specific dataset.

\begin{table}
	\centering
	\caption{Mean 10-fold cross validation accuracy, for synthetic APT data with
			different percentages of atoms removed 
			and $\cN(0,\sigma^2)$ added noise.}\label{tab:results}
	\begin{tabular}{@{}lllll@{}}
			\toprule
			\diagbox[width=6em]{Sparsity}{Std.\\Dev.} & $\sigma = 0$ & $\sigma = 0.25$ & $\sigma = 0.75$ & $\sigma = 1$\\
			\midrule
			0\%  & 100\% & 99.67\% & 98\% & 97.67\% \\ 
			33\% & 100\% &	99.32\% & 96.67\% & 94.67\%\\
			50\% & 97.33\% & 100\% & 92.67\% & 94\%\\
			67\% & 98.67\% & 100\% & 99.33\% & 92\%\\
			\bottomrule
	\end{tabular}
\end{table}

\subsubsection{Imbalanced Classification}\label{sect:vary}

Continuing our study of the binary classification problem, we investigated the effect of
	varying the proportion of BCC vs. FCC lattice structures had on the resulting classification accuracy. We 
	considered the same combinations of sparsity and additive noise as in~\cref{sect:sensitivity},
	but we varied the proportion of BCC structures in the entire dataset between 
	10\% and 90\%. The remaining percentage was composed of FCC structures so that the total
	number of structures was 5,000. We observe a level of accuracy in this 
	setting similar to those observed in the previous experiment; these accuracy scores
	are presented in~\cref{tab:vary_prop}. We observe that the classification scheme is robust
	against not only the perturbations and missing data expected from an APT experiment, but class
	imbalance as well.

	\begin{table}
	\centering
	\caption{Mean 10-fold cross validation accuracy, for synthetic APT data with
			$\cN(0,\sigma^2)$ added noise and 50\% missing, the 
			dataset is comprised of 5,000 configurations in each experiment. 
			The proportion of BCC structures are given, and
			varied between experiments. The proportion of FCC configurations is 1-BCC\%.}\label{tab:vary_prop}
	\begin{tabular}{@{}lllllll@{}}
			\toprule
			BCC proportion & 10\% & 25\% & 40\% & 60\% & 75\% & 90\% \\
			\midrule
			Std. Dev &  \multicolumn{6}{c}{Accuracy}  \\
			\midrule
			{$\sigma = 0.25$} & 96.72\% & 92.32\% & 88.56\% & 88.48\% & 90.24\% & 94.24\% \\
			{$\sigma = 0.5$} & 99.96\% & 99.84\% & 100\% & 100\% & 100\% & 100\% \\
			{$\sigma = 0.75$} & 95.76\% & 89.86\% & 82.88\% & 82.24\% & 85.84\% & 95.04\% \\
			{$\sigma = 1$} & 94.72\% &85.76\% & 81.44\% & 83.2\% & 84.08\% & 94\% \\
			\bottomrule
	\end{tabular}
\end{table}

	\subsubsection{Multi-class Classification}\label{sect:three_way}
	
	As a final experiment, to the previous setting of varying the proportion of BCC vs.\,FCC structures,
		 we 
	add a constant number of HCP structures to the data set. All  
	lattice structures in this experiment are perturbed by Gaussian noise with a standard 
	deviation of 0.25, as the noise
	was found in a previous study to follow a narrowly peaked distribution,
	as opposed to a wide Gaussian distribution~\cite{gault2012atom}. From each of these 
	datasets, we removed $\gamma\%$ of the atoms.
	The results
	of this experiment are in~\cref{tab:vary_prop_all3}. In this scenario, the primary challenge 
	is to correctly identify the FCC and HCP lattices. 
	While these two
	structures are distinct, they have the same density, i.e., the same number of atoms
	per unit volume, and only have a subtle variation in their
	identifying geometry.
	Indeed, there is a
	non-trivial decrease in accuracy when the HCP lattices are introduced into the dataset. Specifically,
	the accuracy declines 
	as the proportion of FCC structures increases relative to the number of HCP lattice types and 
	is the dominant class represented in the dataset. 
	When the BCC proportion comprises 10\% of the
	dataset, the proportion of FCC to HCP lattices is approximately 2:1, and the classifier's 
	accuracy is decreased as compared to settings with less class imbalance in the dataset.

\begin{table}
	\centering
	\caption{Mean 10-fold cross validation accuracy, classifying synthetic APT data with
			$\cN(0,0.25^2)$ added noise and proportion $\gamma\in(0,1)$ missing. We consider three classes, BCC, FCC, and HCP structures,
			in this synthetic APT dataset. We varied the proportion of 
			5,000 configurations between BCC and FCC lattices. The BCC proportion of these structures are given
			and the fraction of FCC configurations is 1-BCC\%.
			To these 5,000 structures we added 
			a constant 2,500 HCP lattice structures in each instance.}\label{tab:vary_prop_all3}
	\begin{tabular}{@{}lllllll@{}}
			\toprule
			BCC proportion & 10\% & 25\% & 40\% & 60\% & 75\% & 90\% \\
			\midrule
			Proportion Missing &  \multicolumn{6}{c}{Accuracy}  \\
			\midrule
			$\gamma=0.33$ & 60.67\% & 69.84\%  & 84.84\% & 86.51\% & 78.88\% & 88.39\%  \\
			$\gamma=0.50$ & 68.33\% & 74.76\%  & 85.16\% & 88.13\% & 82.40\% & 89.45\%  \\
			\bottomrule
	\end{tabular}
\end{table}

\subsection{APT Experimental Data}

We now turn to our original problem of determining the local lattice structure of an HEA from the
experimental APT data. We apply our materials fingerprinting method 
to the APT experimental data from two HEAs, Al$_{1.3}$CoCrCuFeNi and Al$_{0.3}$CoCrFeNi (FCC). 
Recalling \cref{sect:apt_data}, the former has both BCC and 
FCC phases, while the was determined
to be FCC through XRD experiments~\cite{diao2019novel}.
The challenge is to uncover the true atomic-level structure amid the noise and missing data.
 Using our materials fingerprinting methodology, we are able to classify the lattice structure of 200,000 atomic neighborhoods, split evenly between BCC and FCC lattice types, from these APT datasets at \textbf{99.97}\% accuracy with 10-fold cross validation.


\section{Discussion}\label{sect:conclusion}

We have described materials fingerprinting, a topologically-based methodology
for classifying the crystal structure of the HEA APT data
with near-perfect accuracy especially in the binary case.
Starting from a collection of atomic neighborhoods generated by an APT experiment, we 
extract the fundamental topology of the structure and record the information
in a persistence diagram. These diagrams succinctly encode the essential 
topology of an atomic neighborhood
over different length scales in various dimensions. It is by computing the 
persistent homology of the data that we are able to see through the noise
and fill in the sparsity to see where these lattice structures are connected
and where they are not.
Our materials fingerprinting methodology uses the mean and variance of the
$d_p^c$ distance between persistence diagrams to create 
input for a machine learning algorithm. 
This distance not only measures differences in the diagrams
but accounts for different numbers of points between
diagrams being compared. This latter point is salient, as BCC and FCC unit
cells each contain a different number of atoms, and this distinction must be taken into account.  
Basing our materials fingerprint on topological 
features in conjunction with the number of
atoms in each neighborhood, we 
 represent the necessary topological 
and numeric information
required to differentiate between the lattice structures considered here,
with the appropriate choice of metric. 
Indeed, by adopting this point of view, we are
able to qualitatively retain the essential geometric information
of these crystal structures and use this information to predict with greater 
than 99\% accuracy the crystal structure of real APT data.
%

The impact of the present work is two-fold. 
First, the input data to our algorithm is 
point clouds 
generated by HEAs resulting from APT experiments.
The process can be generalized to other lattice types by incorporating
additional crystal structures into the 
materials fingerprint training set. Indeed, the methodology described
	herein does not depend on the labels of the data. It takes in the materials data
	and creates the information-rich persistence diagrams, from which we examine homological
	differences between the diagrams in various dimensions. 
	The data analysis can be performed on multiphase samples,
	although the characterization of individual 
	configurations may need be to be first preceded by classification of domains based on compositional differences, for example. 
	An alternative for comparisons between a multitude of structures is outlined 
	in~\cite{townsend2020representation}, in which different topological descriptors 
	are invoked that consider the electronegativity of 
	the atoms as a feature when creating the persistence diagrams. Such a methodology
	may be used in conjunction with a previous work~\cite{spannausBPSR} that identifies a mapping 
	between the APT data and a known crystal structure, to aid researchers in understanding
	the local structure of materials characterized through the APT process.

\section*{Acknowledgments}
The authors are grateful to
two anonymous referees for helpful comments and suggestions that substantially improved the manuscript.
The APT experiments were conducted at the Oak Ridge National Laboratory's Center for 
Nanophase Materials Sciences (CNMS), which is a U.S. DOE Office of Science User Facility.
The authors would like to thank Jonathan Poplawsky for insightful discussions about the APT method.
V.~M. is grateful for support from 
ARO Grant \# W911NF-17-1-0313 and the NSF DMS-1821241. D.K. and V.M. are grateful for
support from a UTK Seed grant. A.S., C.M., and F.N, acknowledge the Mathematics Department of the University
of Tennessee, where A.S. and C.M. conducted this research as part of their Ph.D studies
and F.N. was a Post-Doctoral research associate. This research used resources of the Compute and Data Environment for Science (CADES) at the Oak Ridge National Laboratory, which is supported by the Office of Science of the U.S. Department of Energy under Contract No. DE-AC05-00OR22725.


\bibliographystyle{acm}
\bibliography{{refs}}

\end{document}